\pdfoutput=1
\documentclass[11pt,twoside,a4paper,cmspaper,final,collab]{cms-tdr}
\def\svnVersion{5e1b48b}\def\svnDate{2026/01/20}\def\cmsCernNoTag{CERN-EP-2025-190}\def\cmsCernDate{\today}\def\cmsMessage{Published in the European Physical Journal C as \href{http://dx.doi.org/10.1140/epjc/s10052-025-15176-9}{\doi{10.1140/epjc/s10052-025-15176-9}.}}
\begin{document}\cmsNoteHeader{B2G-24-015}

\newcommand{\uu}{\ensuremath{\cPqu \cPaqu}\xspace}
\newcommand{\datalumi}{138\fbinv}
\newcommand{\XHY}{\ensuremath{\PX \to \PH \mathrm{Y}}\xspace}
\newcommand{\PY}{\ensuremath{\HepParticle{Y}{}{}}\xspace}
\newcommand{\VV}{\ensuremath{\PV\PV}\xspace}
\newcommand{\MY}{\ensuremath{M_{\PY}}\xspace}
\newcommand{\MX}{\ensuremath{M_{\PX}}\xspace}
\newcommand{\mjj}{\ensuremath{M_\mathrm{jj}}\xspace}
\newcommand{\deta}{\ensuremath{\Delta \eta_{\mathrm{jj}}}\xspace}
\newcommand{\mjy}{\ensuremath{M_\mathrm{j}^{\PY}}\xspace}
\newcommand{\XHYbbWW}{\ensuremath{\XHY \to \bbbar \PWp \PWm \xspace}}
\newcommand{\qqqq}{\ensuremath{2\PQq 2\PAQq^\prime}\xspace}
\newcommand{\ww}{\ensuremath{\PW\PW}\xspace}
\newcommand{\Yww}{\ensuremath{\PY \to \PWp\PWm}\xspace}
\newcommand{\Yuu}{\ensuremath{\PY \to \uu}\xspace}
\newcommand{\Ytt}{\ensuremath{\PY \to \ttbar}\xspace}
\newcommand{\Ybqq}{\ensuremath{\PY \to \PQb \PQq \PAQq^\prime}\xspace}
\newcommand{\XHYbbWWq}{\ensuremath{{\PX\to(\PH\to\bbbar)(\PY\to\ww\to\qqqq)}}\xspace}
\newcommand{\rpf}{\ensuremath{R_{\mathrm{P/F}}}\xspace}
\newcommand{\Hbb}{\ensuremath{\PH\to\bbbar}\xspace}
\newcommand{\PS}{\ensuremath{\HepParticle{S}{}{}}\xspace}
\newcommand{\PSpr}{\ensuremath{\HepParticle{\PS}{}{\prime}}\xspace}
\newcommand{\ppSpHS}{\ensuremath{\Pp\Pp \to \PSpr \to \PH +\PS}\xspace}
\newcommand{\Zjets}{\ensuremath{\PZ\!+\!\text{jets}}\xspace}
\newcommand{\Wjets}{\ensuremath{\PW\!+\!\text{jets}}\xspace}
\newcommand{\sigmaobs}{\ensuremath{\sigma_{\text{obs}}}\xspace}
\newcommand{\sigmaexp}{\ensuremath{\sigma_{\text{bkg}}}\xspace}
\newcommand{\ParticleNet}{\textsc{ParticleNet}\xspace}
\newcommand{\pp}{\ensuremath{\Pp{}\Pp}\xspace}
\newcommand{\com}{\ensuremath{\sqrt{s}=13\TeV}\xspace}
\newcommand{\etaphi}{\ensuremath{\eta-\phi}\xspace}
\newcommand{\rpfmjjmjy}{\ensuremath{\rpf (\mjj,\mjy)}\xspace}
\newcommand{\nFnonres}{\ensuremath{n_{\text{F,nr}}}\xspace}
\newcommand{\absEta}{\ensuremath{\abs{\eta}}\xspace}
\newcommand{\WKK}{\ensuremath{\PW_{\mathrm{KK}}}\xspace}
\ifthenelse{\boolean{cms@external}}{\providecommand{\cmsLeft}{upper\xspace}}{\providecommand{\cmsLeft}{left\xspace}}
\ifthenelse{\boolean{cms@external}}{\providecommand{\cmsRight}{lower\xspace}}{\providecommand{\cmsRight}{right\xspace}}

\cmsNoteHeader{B2G-24-015}
\title{Search for resonances decaying to an anomalous jet and a Higgs boson in proton-proton collisions at \texorpdfstring{\com}{sqrt(s)=13 TeV}}
\date{\today}

\abstract{This paper presents a search for new physics through the process where a massive particle, \PX, decays into a Higgs boson and a second particle, \PY. The Higgs boson subsequently decays into a bottom quark-antiquark pair, which is reconstructed as a single large-radius jet. The decay products of \PY are also assumed to produce a single large-radius jet. The identification of the \PY particle is enhanced by computing the anomaly score of its candidate jet using an autoencoder, which measures deviations from typical quark- or gluon-induced jets. This allows a simultaneous search for multiple \PY decay scenarios within a single analysis. In the main benchmark process, \PY is a scalar particle that decays into a \PW boson pair. Two other scalar \PY decay processes are also considered as benchmarks: decays to a light quark-antiquark pair, and decays to a top quark-antiquark pair. A fourth benchmark process considers \PY as a hadronically decaying top quark, arising from the decay of a vector-like quark into a top quark and a Higgs boson. Data recorded by the CMS experiment at a center-of-mass energy of 13\TeV in 2016--2018, corresponding to an integrated luminosity of \datalumi, are analyzed. The search covers \PX masses between 1.4 and 3.0\TeV and \PY masses between 90 and 400\GeV, with all simulated signals produced in the narrow-width approximation. No significant excess above the standard model background expectation is observed. The most stringent upper limits to date are placed on benchmark signal cross sections for various masses of X and Y particles.}

\hypersetup{
pdfauthor={CMS Collaboration},
pdftitle={Search for resonances decaying to an anomalous jet and a Higgs boson decaying to bbbar in proton-proton collisions at sqrt(s)=13 TeV},
pdfsubject={CMS},
pdfkeywords={LHC, anomaly detection, Higgs boson, merged-jet topology}}

\maketitle 

\section{Introduction}\label{sec:introduction}

The standard model (SM) of particle physics~\cite{Glashow:1961tr,Weinberg:1967tq,Salam:1968rm} is an exceptionally successful theory, demonstrating remarkable agreement across the whole range of experimental observations. Despite its success, there are indications that point toward the existence of physics beyond the SM (BSM). These include, among others, the unknown nature of dark matter~\cite{darkmatter}, and the insufficient violation of the combined charge conjugation and parity symmetry to explain the observed baryon asymmetry of the universe~\cite{cpviolation}.

To address such questions, a multitude of BSM theoretical models, such as supersymmetry~\cite{Fayet:1974pd,Fayet:1977yc}, large or warped extra dimensions~\cite{ArkaniHamed:1998rs,Randall:1999ee,RS-1999}, and those incorporating additional scalar or gauge particles~\cite{Robens:2019kga} have been proposed. A common feature of these theoretical models is the prediction of new particles. Therefore, one avenue of exploration for BSM physics involves resonance searches, which aim to reconstruct the four-momentum of a new particle by analyzing its decay products.
Recent advancements in machine-learning techniques have further motivated the development of inclusive, model-independent searches capable of probing multiple decay scenarios without reliance on specific assumptions~\cite{Kasieczka:2021xcg}. Signatures involving the Higgs boson are well motivated in BSM searches, as new particles may preferentially couple to it~\cite{xhy_review,Domingo2022}.

\begin{figure}[!htb]
  \centering
  \includegraphics[width=0.49\textwidth]{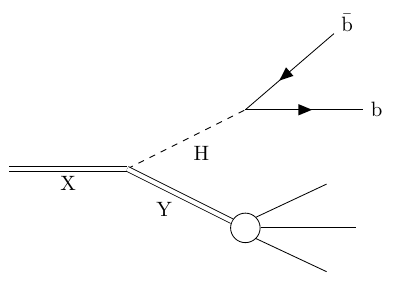}
  \caption{An illustration showing the signal targeted by this analysis. The final state consists of a large-radius jet originating from \PH decaying to \bbbar and another large-radius jet originating from the decay of a second particle, \PY.}
  \label{fig:feyn}
\end{figure}

We search for processes of the form $\PX \to \PH \PY$, where \PX is a heavy BSM resonance, \PH is the 125\GeV Higgs boson, and \PY can be either an SM or a BSM particle. We focus on the case where \PH decays to a bottom quark-antiquark pair (\bbbar), its most probable decay mode, and \PY gives rise to a hadronic jet. The analysis targets the regime where \PX is sufficiently massive that both \PH and \PY are produced with large Lorentz boosts, causing their decay products to be collimated and reconstructed as large-radius jets clustered with a distance parameter of 0.8~\cite{Cacciari:2008gp, Cacciari:2011ma}. An illustration of the process of interest is shown in Fig.~\ref{fig:feyn}. The analysis uses proton-proton (\pp) collision data at \com collected by the CMS experiment at the CERN LHC in 2016--2018, and corresponding to an integrated luminosity of \datalumi. 

We evaluate the results in four benchmark signal scenarios. In the main scenario, \PY is a scalar particle decaying to a \VV pair, where \PV (\PW or \PZ boson) decays hadronically. We choose the \ww decay as the proxy for this process; however, similar sensitivity is expected also for the $\PZ\PZ$ or $\PW\PZ$ decays. The decay allows for one or both vector bosons to be off-shell.
Two additional scalar particle decay modes are considered, with \PY decaying either into a light-flavored quark-antiquark pair (\qqbar) or a top quark-antiquark pair (\ttbar), in which both top quarks decay fully hadronically. In the fourth scenario, \PY is a top quark decaying fully hadronically. These four scenarios span a range of jet substructures, enabling the evaluation of the analysis performance across diverse signal signatures. The first three benchmark scenarios are motivated by theoretical models, such as the next-to-minimal supersymmetric extension of the SM (NMSSM)~\cite{EllwangerNMSSM1}, in which cascade decays involving more than one new particle become possible. Reference~\cite{xhy_review} provides an overview of CMS searches for the \ppSpHS process, where $\mathrm{S'}$ and \PS represent new scalar particles and \PS is assumed to decay to \bbbar. However, depending on the NMSSM parameters, other \PS decay modes may be favored, motivating searches targeting alternative \PS decays. The presented analysis thus complements Ref.~\cite{xhy_review} by exploring benchmark models with different \PS decay modes.
The fourth signal scenario is motivated by models predicting the existence of vector-like quarks~\cite{Carvalho_2018}.

This work builds on a generic dijet anomaly search performed by the CMS Collaboration~\cite{case}, but it specifies one particle and its decay mode (\Hbb). This enables the application of more specific event selection criteria, enhancing the sensitivity of the analysis, while retaining a degree of generality in the characterization of the second particle.

The \ParticleNet~\cite{Qu:2019gqs} jet tagging algorithm, is applied to \PH candidate jets to assign scores indicating consistency with the decay of a massive resonance into \bbbar. The \ParticleNet score is used to define ``Pass'' and ``Fail'' analysis regions.
An autoencoder (AE)~\cite{case} is applied to \PY candidate jets to assign anomaly scores, with higher scores indicating deviation from jets typical of multijet production via quantum chromodynamics (QCD). This approach enables the identification of jets with unusual substructure, potentially arising from BSM decays, without relying on specific signal models. The anomaly score is used to define measurement (MR) and control (CR) analysis regions. Thus, the analysis uses four orthogonal regions, with the search performed in MR Pass. The specific selection criteria for these regions are defined in Section~\ref{sec:EvtSel}.

The search is carried out in a two-dimensional (2D) plane defined by the mass of the \PY candidate jet, \mjy, and the dijet invariant mass of the \PY and \PH candidate jets, \mjj. For a signal process, the two variables correspond to the masses of \PY and \PX, denoted as \MY and \MX, respectively. As a result, the analysis searches for a localized excess of events in the \mjj-\mjy plane.

The paper is organized as follows. After the description of the CMS detector and event reconstruction in Section~\ref{sec:detector}, we describe signal and background processes in Section~\ref{sec:Simulation}. The event selection is discussed in Section~\ref{sec:EvtSel}, followed by the description of the statistical model in Section~\ref{sec:statModel} and systematic uncertainties in Section~\ref{sec:systematic}. The results are presented in Section~\ref{sec:Results}, followed by a summary of the paper in Section~\ref{sec:Summary}.

\section{The CMS detector and event reconstruction}\label{sec:detector}
The CMS apparatus~\cite{CMS:2008xjf} is a multipurpose, nearly hermetic detector, designed to trigger on~\cite{CMS:2020cmk,CMS:2016ngn,CMS:2024psu} and identify electrons, muons, photons, and charged and neutral hadrons~\cite{CMS:2020uim,CMS:2018rym,CMS:2014pgm}. Its central feature is a superconducting solenoid of 6\unit{m} internal diameter, providing a magnetic field of 3.8\unit{T}. Within the solenoid volume are a silicon pixel and strip tracker, a lead tungstate crystal electromagnetic calorimeter (ECAL), and a brass and scintillator hadron calorimeter (HCAL), each composed of a barrel and two endcap sections. Forward calorimeters extend the pseudorapidity ($\eta$) coverage provided by the barrel and endcap detectors. Muons are reconstructed using gas-ionization detectors embedded in the steel flux-return yoke outside the solenoid. More detailed descriptions of the CMS detector, together with a definition of the coordinate system used and the relevant kinematic variables, can be found in Refs.~\cite{CMS:2008xjf,CMS:2023gfb}. 

Events of interest are selected using a two-tiered trigger system. The first level, composed of custom hardware processors, uses information from the calorimeters and muon detectors to select events at a rate of around 100\unit{kHz} within a fixed latency of about 4\mus~\cite{CMS:2020cmk}. The second level, known as the high-level trigger, consists of a farm of processors running a version of the full event reconstruction software optimized for fast processing, and reduces the event rate to a few kHz before data storage~\cite{CMS:2016ngn,CMS:2024psu}. 

A particle-flow algorithm~\cite{CMS:2017yfk} aims to reconstruct and identify each individual particle in an event, with an optimized combination of information from the various elements of the CMS detector. The energy of photons is obtained from the ECAL measurement. The energy of electrons is determined from a combination of the electron momentum at the primary interaction vertex as determined by the tracker, the energy of the corresponding ECAL cluster, and the energy sum of all bremsstrahlung photons spatially compatible with originating from the electron track. The energy of muons is obtained from the curvature of the corresponding track. The energy of charged hadrons is determined from a combination of their momentum measured in the tracker and the matching ECAL and HCAL energy deposits, corrected for the response function of the calorimeters to hadronic showers. Finally, the energy of neutral hadrons is obtained from the corresponding corrected ECAL and HCAL energies. 

For each event, hadronic jets are clustered from these reconstructed particles using the infrared- and collinear-safe anti-\kt algorithm~\cite{Cacciari:2008gp, Cacciari:2011ma} with a radius parameter of either 0.4 (AK4 jets) or 0.8 (AK8 jets). Jet momentum is determined as the vectorial sum of all particle momenta in the jet, and is found from simulation to be, on average, within 5--10\% of the true momentum over the whole transverse momentum (\pt) spectrum and detector acceptance. Additional \pp interactions within the same or nearby bunch crossings (pileup) can contribute additional tracks and calorimetric energy depositions to the jet momentum. The pileup-per-particle identification algorithm~\cite{Sirunyan:2020foa,Bertolini:2014bba} is used to mitigate the effect of pileup at the reconstructed particle level, making use of local shape information, event pileup properties, and tracking information. A local shape variable is defined, which distinguishes between collinear and soft diffuse distributions of other particles surrounding the particle under consideration. The former is attributed to particles originating from the hard scatter and the latter to particles originating from pileup interactions. Charged particles identified to be originating from pileup vertices are discarded. For each neutral particle, this variable is computed using the surrounding charged particles compatible with the primary interaction vertex within the tracker acceptance ($\absEta < 2.5$), and using both charged and neutral particles in the region outside of the tracker coverage. The momenta of the neutral particles are then rescaled according to their probability to originate from the primary interaction vertex deduced from the local shape variable, superseding the need for jet-based pileup corrections~\cite{Sirunyan:2020foa}. 

For AK8 jets, masses are computed after applying grooming~\cite{SalamJetography} techniques, which remove soft, wide-angle and collinear radiation from the jets, in order to mitigate the effects of contamination from initial-state radiation, the underlying event, and multiple hadron scattering. 
The trimming algorithm~\cite{Krohn:2009th} uses a subjet size parameter of 0.3 and a radiation fraction parameter $z=0.1$, which determines the minimum \pt fraction that the reclustered jet constituents need to have in order not to be removed. The mass of the resultant jet is referred to as its ``trimmed mass''.
The ``soft-drop mass'' of the jet is obtained as follows. The constituents of the AK8 jets are reclustered using the Cambridge--Aachen algorithm~\cite{Dokshitzer:1997in,Wobisch:1998wt}. Then, the ``modified mass drop tagger'' algorithm~\cite{Dasgupta:2013ihk,Butterworth:2008iy}, also known as the ``soft drop'' algorithm, is applied. The algorithm is applied with angular exponent $\beta = 0$, soft cutoff threshold $z_{\mathrm{cut}} < 0.1$, and characteristic radius $R_{0} = 0.8$~\cite{Larkoski:2014wba}, which corresponds to the AK8 jet clustering distance parameter.

\section{Signal and background processes\label{sec:Simulation}}

The main benchmark decay mode in this paper is \XHYbbWW, where \PX and \PY are scalar particles, and where both \PW bosons decay hadronically. It is simulated at leading order (LO) using the {\MGvATNLO\xspace 2.6.5}~\cite{Madgraph} event generator and the NMSSM model~\cite{EllwangerNMSSM1,Djouadi_2008}. The samples are generated for 42 different combinations of \MX and \MY values, with \MX ranging from 1400 to 3000\GeV, and \MY ranging from 90 to 400\GeV. The range of \MX is chosen to ensure full trigger efficiency at low mass and sufficient background event yields for the background estimation procedure at high mass, while the \MY values are selected to maintain validity of the Lorentz-boosted regime across the \MX spectrum. Two other scalar decay modes, $\PY\to\qqbar$ (with \uu used as a proxy) and $\PY\to\ttbar$, are also simulated at LO with the NMSSM model using \MGvATNLO 2.6.5, each with a fixed \MY (200 and 400\GeVns{}, respectively) and the same range of \MX values as the main benchmark. The above scalar benchmark processes are modeled assuming production via gluon-gluon fusion.
The \Ybqq decay mode is also considered, motivated by models predicting vector-like quarks~\cite{Carvalho_2018}. In this scenario, \PX is identified as a vector-like quark and \PY as the top quark. Here, the anomalous jet originates from the fully hadronic decay of the top quark, providing an example that such jets do not necessarily have to come from a BSM particle. This scenario is simulated at LO for six different \MX values with \MY set to the mass of the top quark. In the rest of the paper, we differentiate between the signal models by specifying the decay mode of \PY. All generated signal samples assume the narrow-width approximation for \PX and \PY.

The SM background consists of \ttbar production, QCD multijet events, \Wjets and \Zjets processes, and \PH production, with and without an associated vector boson.

The \ttbar events in all-hadronic or lepton(s)+jets final states are modeled using {\POWHEG 2.0}~\cite{POWHEG,POWHEG_Frixione,POWHEGBOX,POWHEG_Nason}, at next-to-LO (NLO). The simulated \ttbar event yields are normalized using a cross section of 834\unit{pb}, calculated at next-to-NLO (NNLO) in QCD with soft gluon resummation at next-to-next-to-leading logarithmic precision~\cite{Czakon:2011xx}. The theoretical uncertainty on this cross section is approximately 5\%, combining scale, parton distribution function (PDF), and top mass uncertainties in quadrature.

  {\tolerance=3000000
    The QCD multijet, \Wjets, and \Zjets event samples are simulated at LO using the \MGvATNLO event generator. These
    backgrounds have nonresonant distributions in the analysis observables and are thus jointly estimated using data from the sidebands. Their simulated samples are only used to develop the tools for the analysis.
    The gluon-gluon fusion \PH boson production process is simulated using the \textsc{MiNLO} event generator~\cite{NNLOPS}. The \POWHEG generator~\cite{Nason:2009ai,Luisoni:2013kna} is used to model \PH production via vector boson fusion and associated production with a \PW or \PZ boson at NLO accuracy.
    \par}

All samples are generated using the NNPDF3.1~\cite{NNPDF31} NNLO PDFs from the {\textsc{lhapdf}{}6} PDF library~\cite{Buckley:2014ana}. The showering and hadronization of partons are simulated with \PYTHIA 8.240~\cite{Pythia82} and the \textsc{CP5} underlying event tune~\cite{CP5}.

All generated events are processed through a simulation of the CMS detector based on \GEANTfour\hspace{0pt}~\cite{GEANT}. The effects of pileup are modeled assuming a total inelastic \pp cross section of {69.2\unit{mb}}~\cite{CMS:2018mlc} by overlapping simulated minimum bias events with the hard-scattering process. All simulated event samples are weighted to match the distribution of pileup vertices observed in the data.

\section{Event selection\label{sec:EvtSel}}

The online selection uses triggers based on significant jet activity. Different trigger thresholds were used in 2016 (2017 and 2018) as follows. One trigger criterion required a single AK8 jet with $\pt>450$ (500)\GeV. A second trigger required that \HT, the scalar sum of the \pt of all AK4 jets with $\pt>30\GeV$ and $\absEta<2.5$, must be greater than 900 (1050)\GeV. Additionally, in 2017--2018, a third trigger algorithm required an AK8 jet with a trimmed mass $>30$\GeV along with $\pt>400\GeV$. The combined logical OR of all the triggers is used. The trigger selection is fully efficient with respect to the offline selection.

The offline selection requires the two leading (in \pt) AK8 jets to have $\pt>300\GeV$ and $|\eta| < 2.4$. A requirement is also placed on the invariant mass of the two jets, $\mjj > 1300\GeV$, where $\mjj$ is computed from the jet four-vectors, including the jet mass. A selection is also placed on the difference in $\eta$ between the jets, $|\deta|<1.3$, in order to further suppress the QCD multijet background, which is characterized by larger jet separation. Finally, the \PH candidate jet, defined below, is required to have its soft-drop mass in the 100--150\GeV range.

  {\tolerance=800
    The \ParticleNet algorithm~\cite{CMS-DP-2020-002} is used to discriminate AK8 jets consistent with the decay of a massive particle into \bbbar against the background from jets originating from QCD multijet production. A mass-decorrelated version of \ParticleNet~\cite{CMS-DP-2022-005} is used to ensure that background-enriched regions can be defined using jets with low \ParticleNet scores while maintaining the same background mass distributions in the regions passing the \ParticleNet selection. The jet with the higher \ParticleNet score is considered to be the \PH candidate jet and the other to be the \PY candidate jet. The soft-drop mass of the \PY candidate jet (\mjy) is used as a second observable alongside \mjj.
    \par}

An AE neural network~\cite{Hinton:2006} is used to quantify the anomaly score of the \PY candidate jet. An AE consists of two parts. The first part, the ``encoder'', compresses an input into a latent representation of reduced dimensionality. The second part, the ``decoder'', reverses this process and attempts to reconstruct the original input from the reduced representation. 
We choose to represent jets as 32$\times$32 pixel images~\cite{image_paper}, covering a region of $\eta$--$\phi$ space with width $\Delta\eta=\Delta\phi =1.2$ centered on the jet axis, where $\phi$ is the azimuthal angle in radians. The intensity of each pixel is determined by the sum of energies from the constituent particles within the given $\eta$--$\phi$ region. We define the anomaly score as the mean squared difference between the pixel intensities of the input image and the reconstructed output. This AE architecture was previously used in Ref.~\cite{case} as part of the ``Tag N' Train'' method (as discussed in Appendix A.2 of Ref.~\cite{case}). 
The training is performed on simulated QCD jets, allowing the AE to compress these jets with minimal information loss. However, for non-QCD jets (from signal, \ttbar production, etc.), which exhibit different substructure patterns, this procedure leads to poor compression and reconstruction of the original jet. This enables effective discrimination of non-QCD jets.

Each event is categorized into one of four groups based on the \ParticleNet score of the \PH candidate jet (pass or fail) and the anomaly score of the \PY candidate jet (control region or measurement region): CR Pass, CR Fail, MR Pass, and MR Fail. The CR is used to validate the background estimation method, described below, whereas the MR is used to extract the signal. The Fail regions are used to help estimate the shape of the QCD multijet background in their corresponding Pass regions.

A \ParticleNet working point corresponding to a ${\approx}60\%$ signal efficiency (defined as the fraction of signal jets passing the selection), as measured on the main benchmark with $\MX = 2000\GeV$ and $\MY = 250\GeV$, and a ${\approx}0.5\%$ mistag rate (defined as the fraction of multijet background jets passing the selection) defines the boundary between the Pass and Fail tagging categories. 

An anomaly score working point with ${\approx}85\%$ signal efficiency and 30\% mistag rate on the same benchmark defines the MR region. The CR consists of jets that fail the anomaly score MR working point. A lower bound on the anomaly score is applied so that the CR remains similar to the MR, which helps with the validation of the background estimation method. As a result, about 10\% of signal and 30\% of QCD jets are assigned to the CR.

The \ParticleNet efficiency for correctly identifying \Hbb jets is calibrated in observed data using a sample of jets originating from fragmentation of a gluon to \bbbar, which are similar to \Hbb jets. Such jets are selected from the observed data using a boosted decision tree classifier, such that their ensemble \ParticleNet score resembles that of \Hbb jets~\cite{CMS-PAS-BTV-22-001}. A systematic uncertainty is assigned to account for the sensitivity of the results to changes in the boosted decision tree requirement applied to the jets. The data-to-simulation correction factors for the \ParticleNet signal selection efficiency range from 0.93--1.05, with uncertainties up to 15\%, depending on the jet \pt and data-taking period. The \ParticleNet efficiency for misidentifying jets from the \ttbar background as \Hbb jets is constrained directly in the fit to data by allowing the yield in the Pass and Fail regions to vary while keeping the total yield constant.

The data-to-simulation scale factors for the efficiency of the anomaly score selection are calculated using the Lund plane reweighting method~\cite{CMS:LundRwt}, separately for each signal process and for the \ttbar background. The method involves applying per-jet event weights based on their distribution in the Lund plane, a 2D representation defined by the logarithms of splitting angle and \pt. These weights are used to adjust the simulated radiation pattern so that the density in the Lund plane matches that observed in the data. It was found that the shapes of the two observables, \mjy and \mjj, do not change significantly with the application of per-jet weights. This allows applying an overall correction to the anomaly tagging efficiency instead of per-jet event reweighting. The corrections are calculated as the ratio of the anomaly tagging efficiency before and after per-jet reweighting. They are found to be compatible with unity, with 10--20\% uncertainty.

\section{Statistical model}\label{sec:statModel}
The analysis searches for a localized signal in the 2D \mjj-\mjy plane. The \mjj (\mjy) axis spans the 1300--3000 (40--500)\GeV range. For higher masses, the bin widths are increased in order to ensure sufficient population of each bin. As a result, all the bins in the MR Fail have 8 or more observed events. This helps to reduce the per-bin statistical uncertainties in the QCD multijet background in the Pass region. The \mjj-\mjy distributions for the background and signal predictions are simultaneously fitted to data in either the MR Pass and MR Fail, or CR Pass and CR Fail. The contributions from different data-taking periods are added together in data and simulation to construct 2D histogram templates. The templates for the \ttbar and \PH production background processes, and for the signal benchmarks, are obtained from simulation. A pass-to-fail ratio method, described in the following paragraphs, is used to model the nonresonant background.

The pass-to-fail ratio method is based on the ratio of \mjj-\mjy distributions between \ParticleNet Pass and Fail regions: \rpfmjjmjy. It is defined through the relation:
\begin{equation}
  n_{\mathrm{P,nr}}(i) = \nFnonres(i)\rpfmjjmjy,
\end{equation}
where $n_{\mathrm{F(P),nr}}(i)$ is the number of nonresonant events in bin $i$ of the \mjj-\mjy distribution in the Fail (Pass) region. Since the Fail region is dominated by nonresonant events, $\nFnonres(i)$ can be directly estimated from data by subtracting the relatively small simulated yields of other processes. The \rpf is a priori unknown and is determined during the fit of signal and background distributions to the data. It is modeled as a polynomial in \mjj and \mjy of order $n$ and is defined as a linear combination of $\mjj^a \mathrm{M}_\mathrm{jY}^b $ terms where the exponents $a$ and $b$ satisfy $a+b\le n$. Simulation shows no strong dependence of the \ParticleNet mistag rate in either of the two observables, whereas the mistag rate of anomaly tagging is observed to rise with jet mass in the simulation. This makes \ParticleNet tagging the preferred method for defining the Pass and Fail regions for background estimation, ensuring that polynomials with low $n$ will be able to well describe {\rpf}. A Fisher's F-test~\cite{ftest} is used to determine the minimum polynomial order necessary and sufficient for the model. Starting from polynomials of order zero, the F-test is used to determine if the next higher order provides a significant improvement in the fit quality. The \rpf polynomial orders in the CR and MR are discussed in Section~\ref{sec:Results}.

Systematic uncertainties, described in Section~\ref{sec:systematic}, are included in the statistical model by generating template histograms representing one standard deviation variations from the nominal simulation. These variations correspond to individual sources of uncertainty. Each is assigned a nuisance parameter in the likelihood function used for statistical interpretation, using a unit Gaussian distribution for shape uncertainties, and a log-normal distribution for normalization uncertainties~\cite{CMS:2024onh}. These distributions change the simulated template shape according to the corresponding template uncertainty histograms.

\section{Systematic uncertainties}\label{sec:systematic}

Several sources of systematic uncertainty are considered, affecting both the shapes and yields of the signal and background templates.
Data-to-simulation scale factors for the \ParticleNet signal and \PH boson background efficiency are calculated with the method discussed in Section~\ref{sec:EvtSel}. They have approximately 10\% uncertainty. 
For the \ttbar background, the mistag rate correction is directly determined during the fit using the Pass and Fail regions. 
The scale factors for signal and \ttbar anomaly tagging efficiency, also discussed in Section~\ref{sec:EvtSel}, have uncertainties of about 15--30\%. For the \PH boson background, a 20\% uncertainty is assigned to the anomaly tagging efficiency.

Initial- and final-state radiation uncertainties are evaluated separately by varying the renormalization scale used in the parton shower by factors of 0.5 and 2.
Measured uncertainties on jet energy scale and resolution are propagated to the jets used in the analysis in order to evaluate their effects. 
Scale and resolution uncertainties are considered uncorrelated. 

The PDF uncertainties are evaluated as the square root of differences of squared PDF Hessian eigenvector sets using NNPDF3.1 PDF sets. The uncertainties coming from the choice of renormalization and factorization scales are evaluated by separately varying each of the scales while keeping the other fixed to the nominal value, and also by scaling both by factors of 0.5 and 2.0. The effect of these variations is found to be negligible for signal acceptance and thus only the \ttbar and \PH boson backgrounds are affected. 

A 1.6\% uncertainty in the total integrated luminosity is applied~\cite{CMS-LUM-17-003,CMS-PAS-LUM-17-004,CMS-PAS-LUM-18-002}. During the 2016--2017 data taking, a gradual shift in the timing of the inputs of the ECAL hardware level trigger in the region of $\absEta > 2$ caused a specific trigger inefficiency. Event-level weight corrections and their uncertainties are included in the simulation to account for this. 
A pileup correction is applied to simulated events based on the true number of interactions in the event. The corresponding uncertainty is generated by varying the inelastic \pp cross section at 13\TeV (69.2\unit{mb}) by 4.6\%~\cite{CMS:2018mlc}. Finally, a 5\% normalization uncertainty is applied to the \ttbar cross section.

The impact of each uncertainty source on the fitted signal strength parameter is estimated by varying the corresponding nuisance parameter within its postfit uncertainty and refitting the signal strength, using the $\MX=2200\GeV$ and $\MY=250\GeV$ simulated signal sample as a representative test case. The uncertainties with the largest impacts, ranging from 5\% to 8\%, are associated with the efficiency of tagging anomalous jets, the efficiency of tagging \Hbb jets, parton showering, and the renormalization and factorization scales. Other uncertainties, such as those related to the jet energy scale and resolution, integrated luminosity, and pileup, have smaller effects, altering the signal strength by 3\% or less.

Overall, the sensitivity of the analysis is statistically limited, due to the limited amount of the observed data.

\section{Results}
\label{sec:Results}

\begin{figure*}[!htb]
  \centering
  \includegraphics[width=0.49\textwidth]{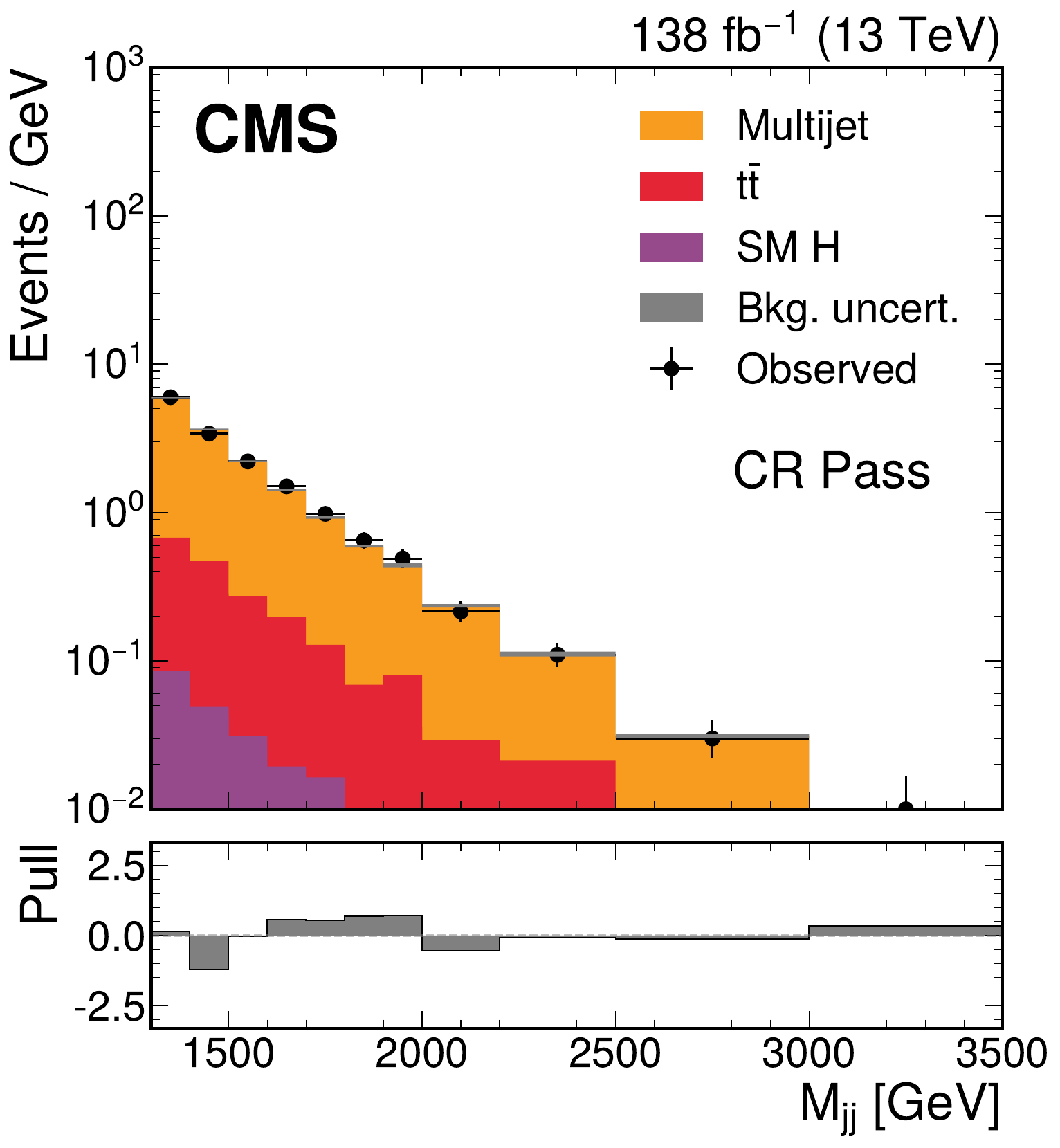}
  \includegraphics[width=0.49\textwidth]{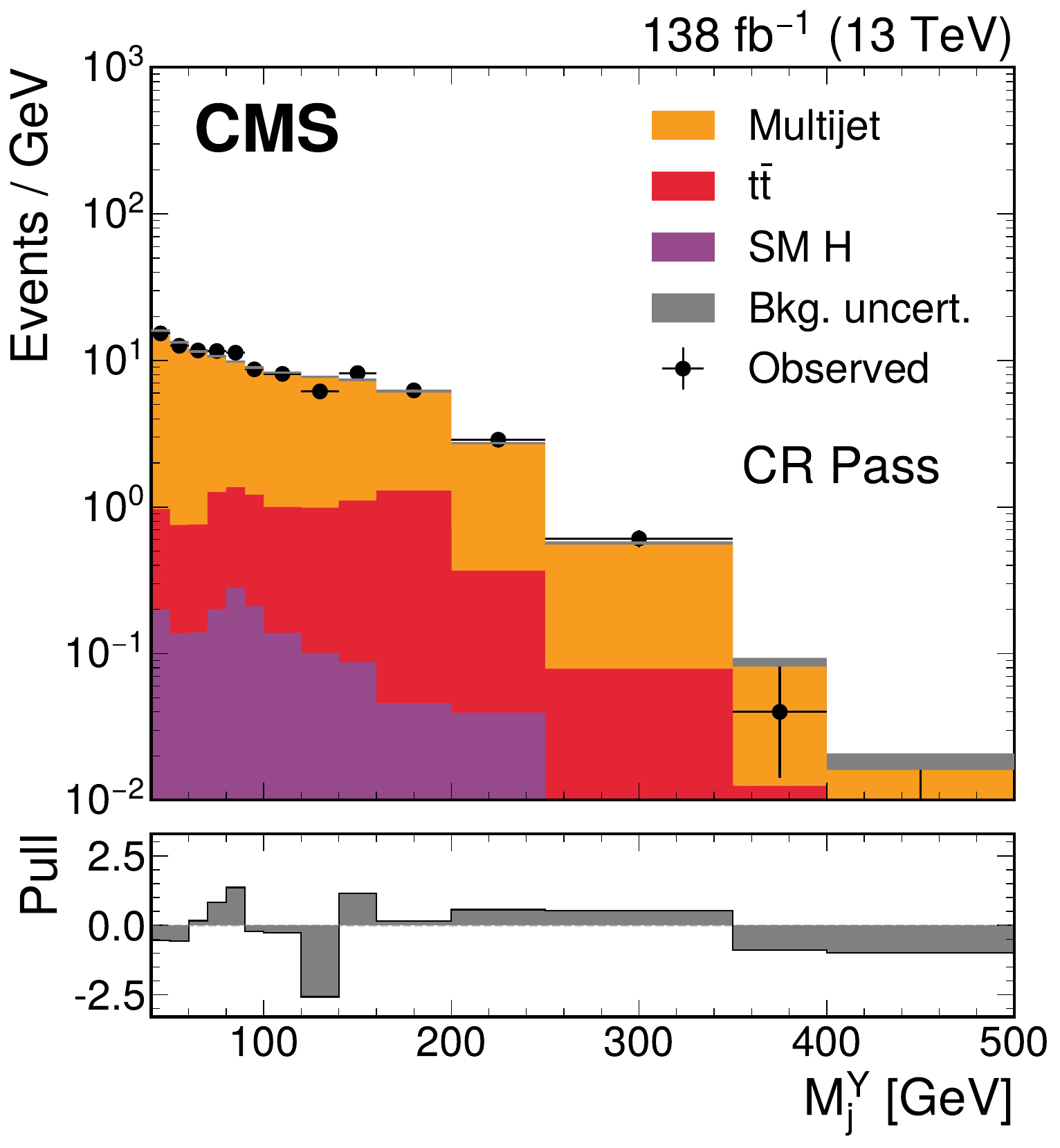}\\
  \includegraphics[width=0.49\textwidth]{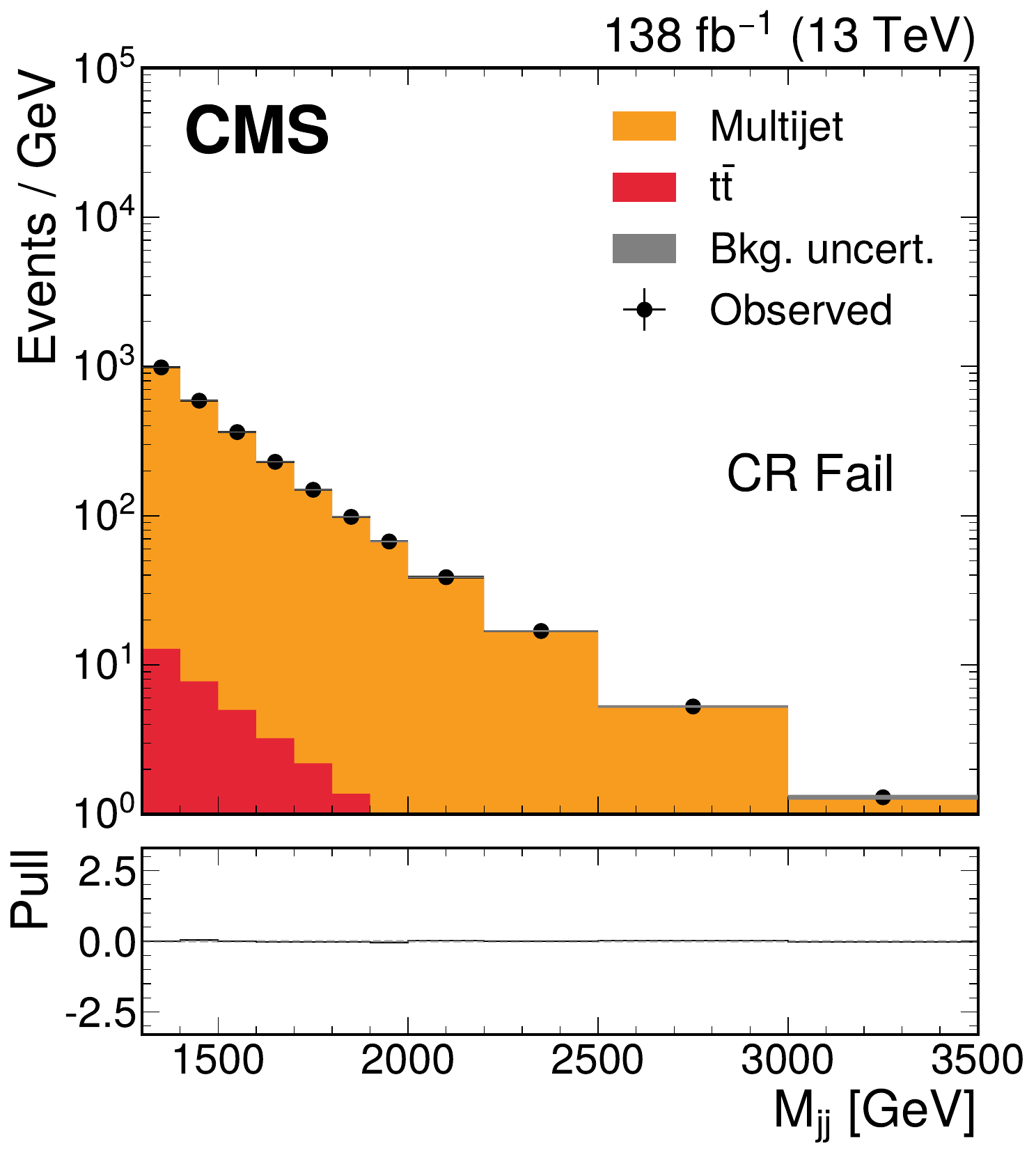}
  \includegraphics[width=0.49\textwidth]{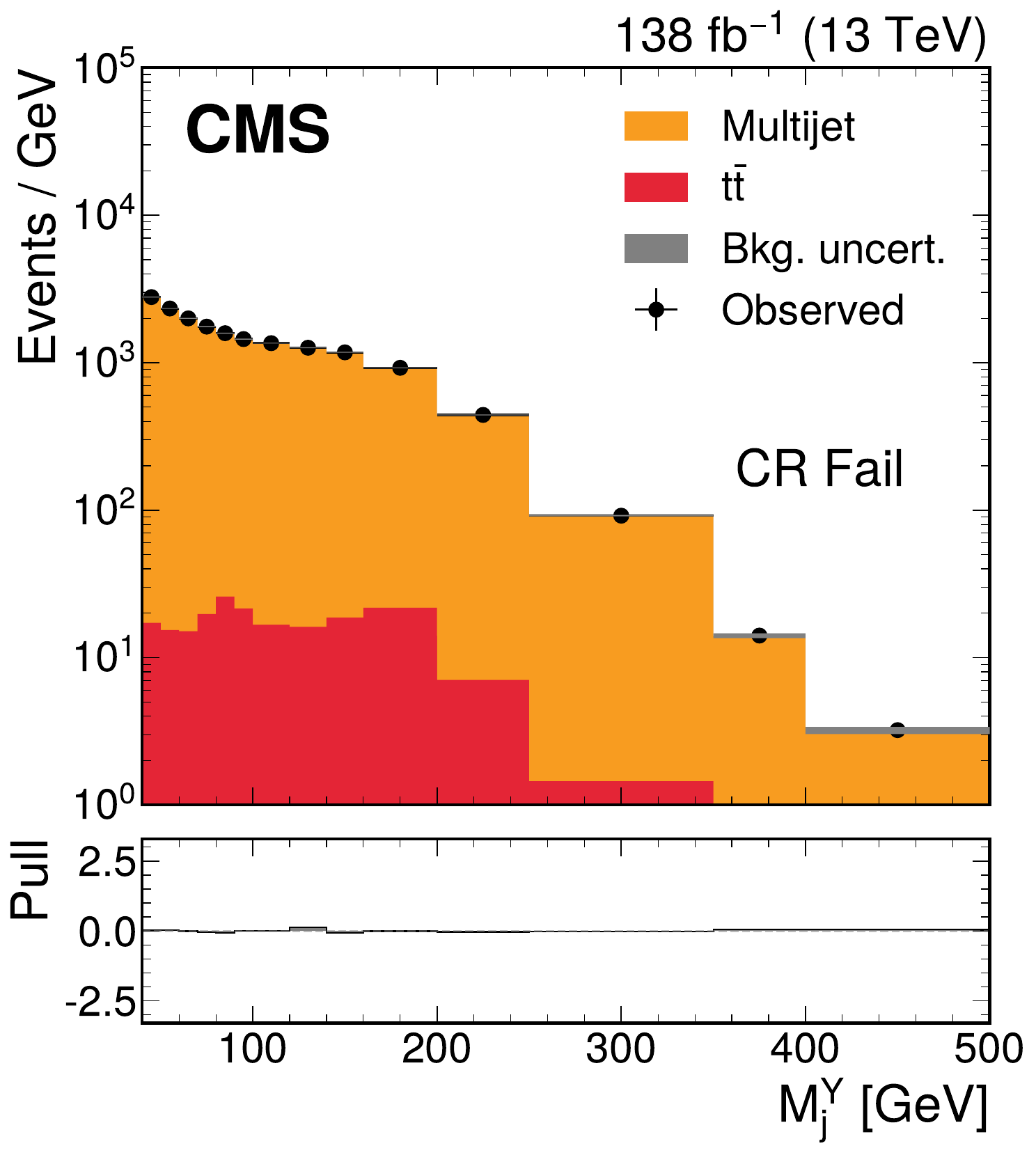}
  \caption{The \mjj (left) and \mjy (right) projections showing the number of observed events per \GeV (black markers) compared with the backgrounds estimated in the fit to the data (filled histograms) in the CR. Pass (upper) and Fail (lower) categories are shown. The high level of agreement between the model and the data in the Fail region arises because this region is used to constrain the background estimate. The lower panels show the ``Pull'' defined as $(\text{observed events}{-}\text{expected events})/\sqrt{\smash[b]{\sigmaobs^{2} + \sigmaexp^{2}}}$, where \sigmaobs and \sigmaexp are the total uncertainties in the observation and the background estimation, respectively.}
  \label{fig:postfit_CR}
\end{figure*}

\begin{figure*}[!htb]
  \centering
  \includegraphics[width=0.49\textwidth]{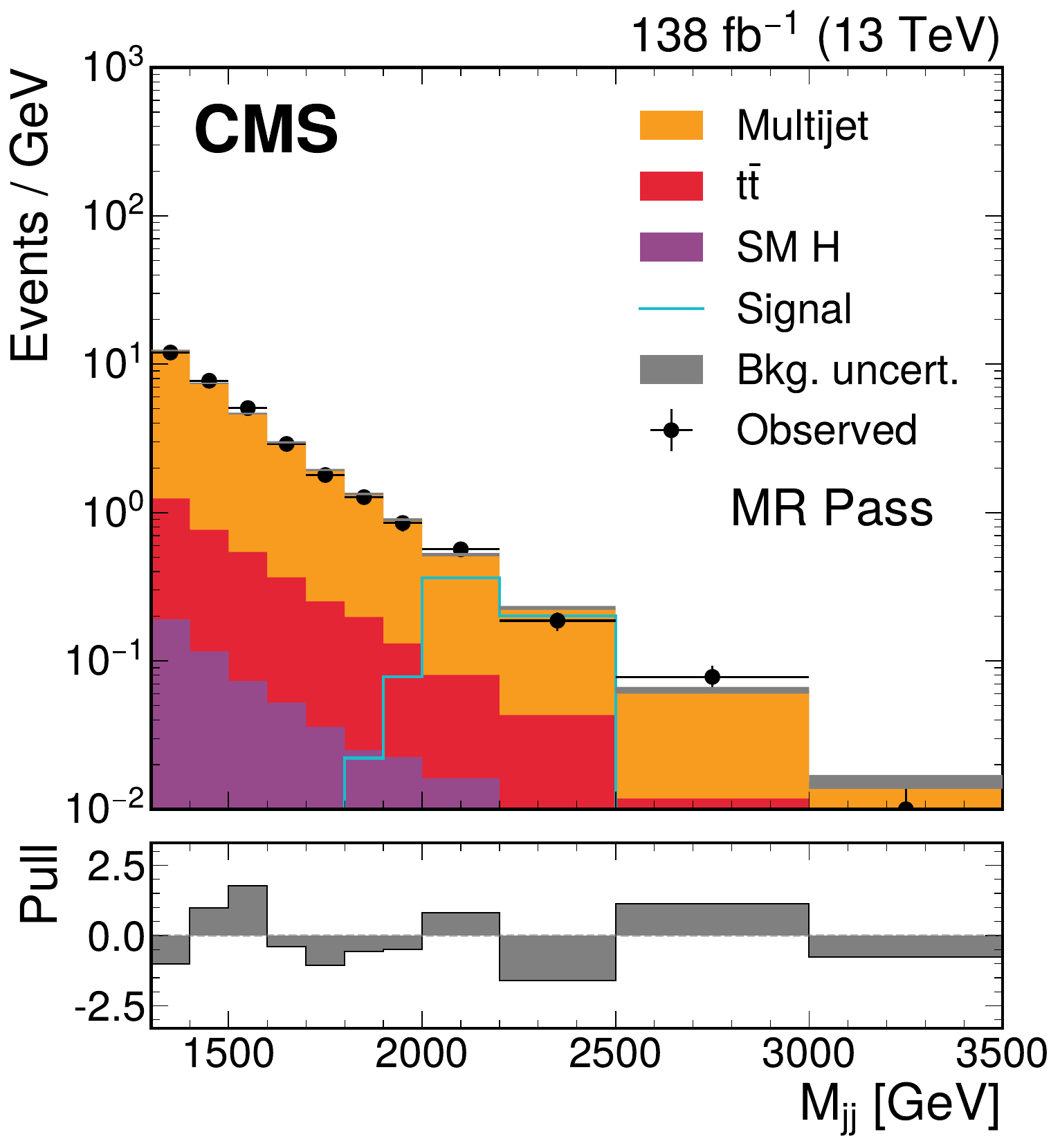}
  \includegraphics[width=0.49\textwidth]{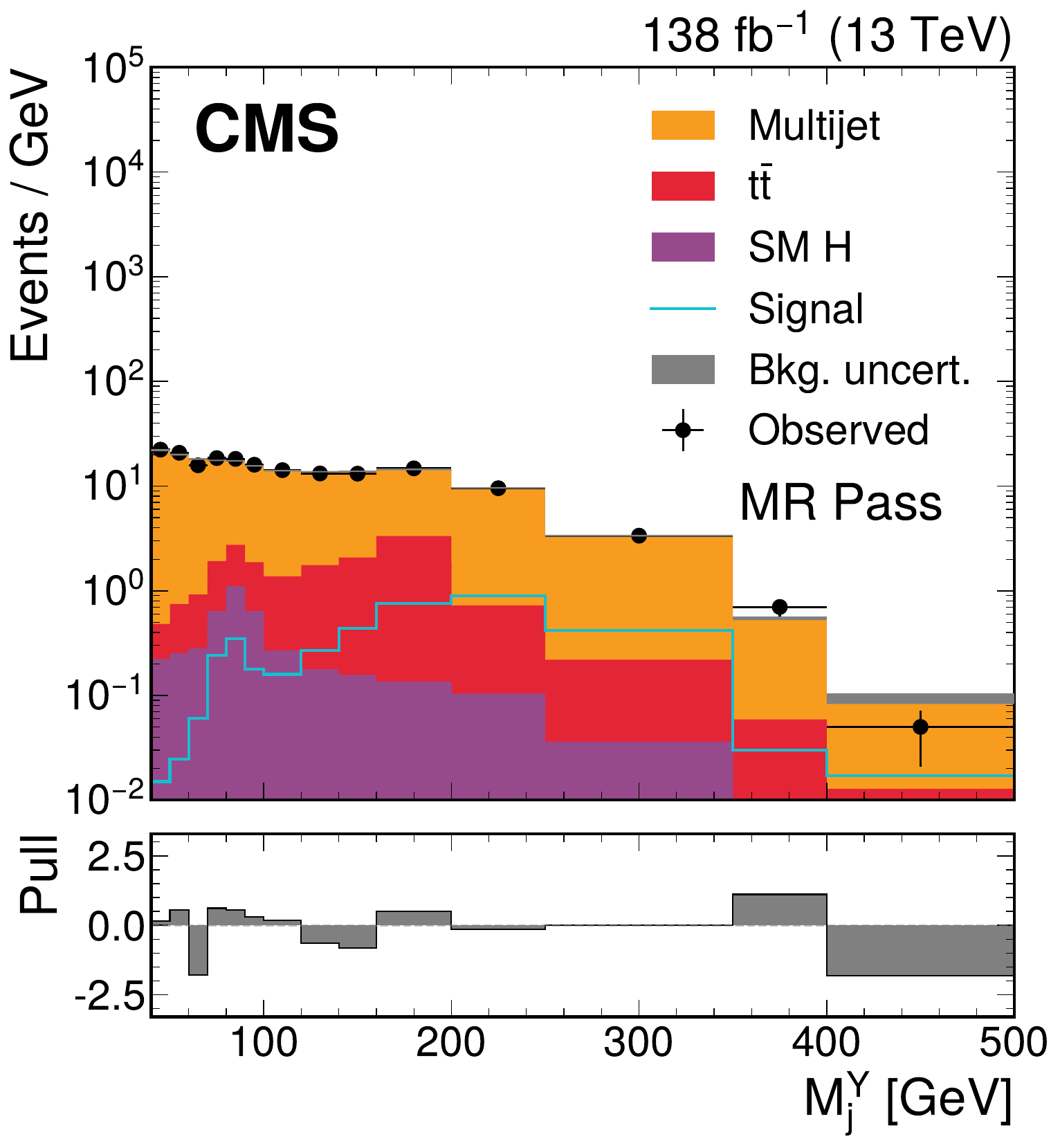}\\
  \includegraphics[width=0.49\textwidth]{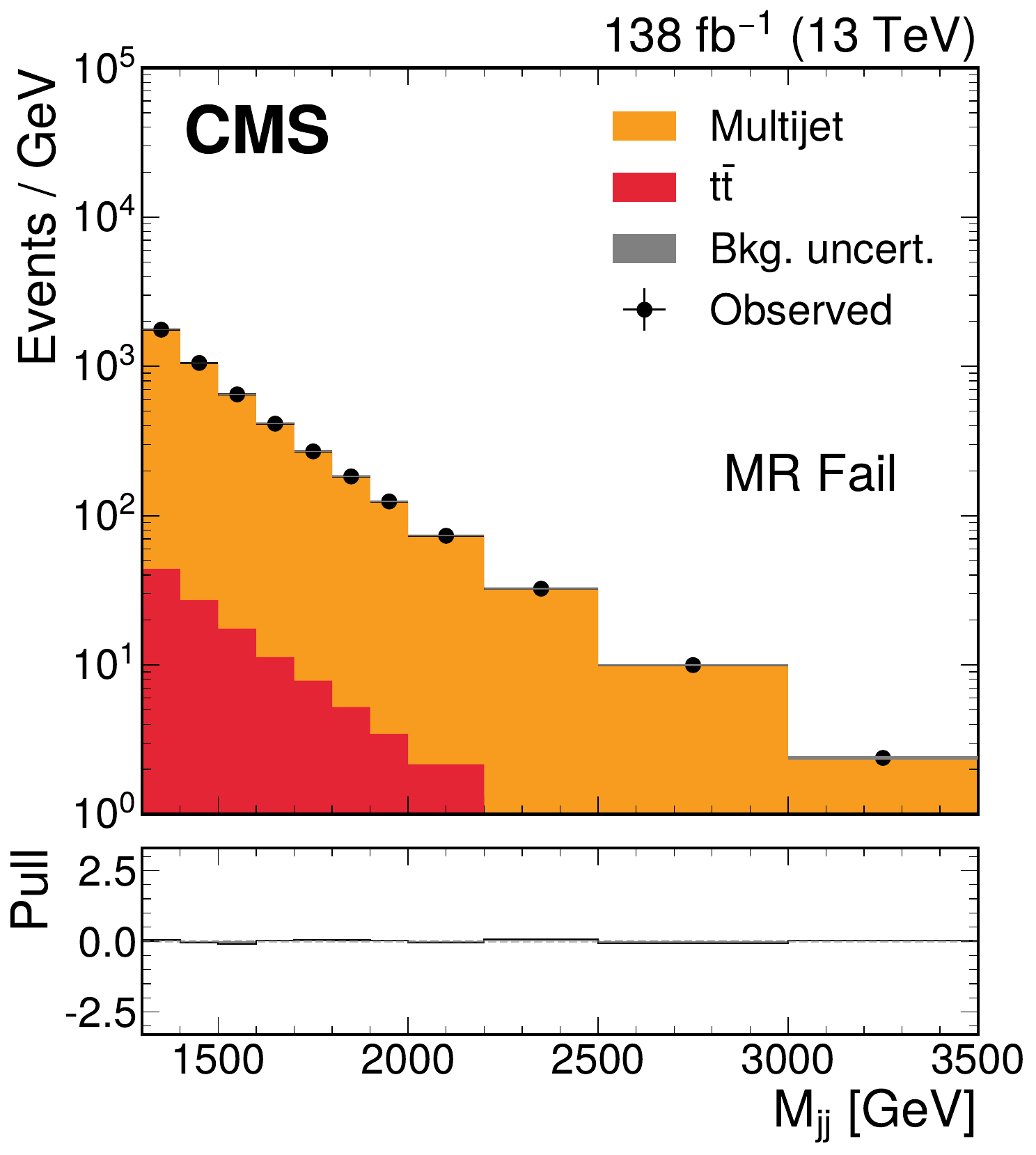}
  \includegraphics[width=0.49\textwidth]{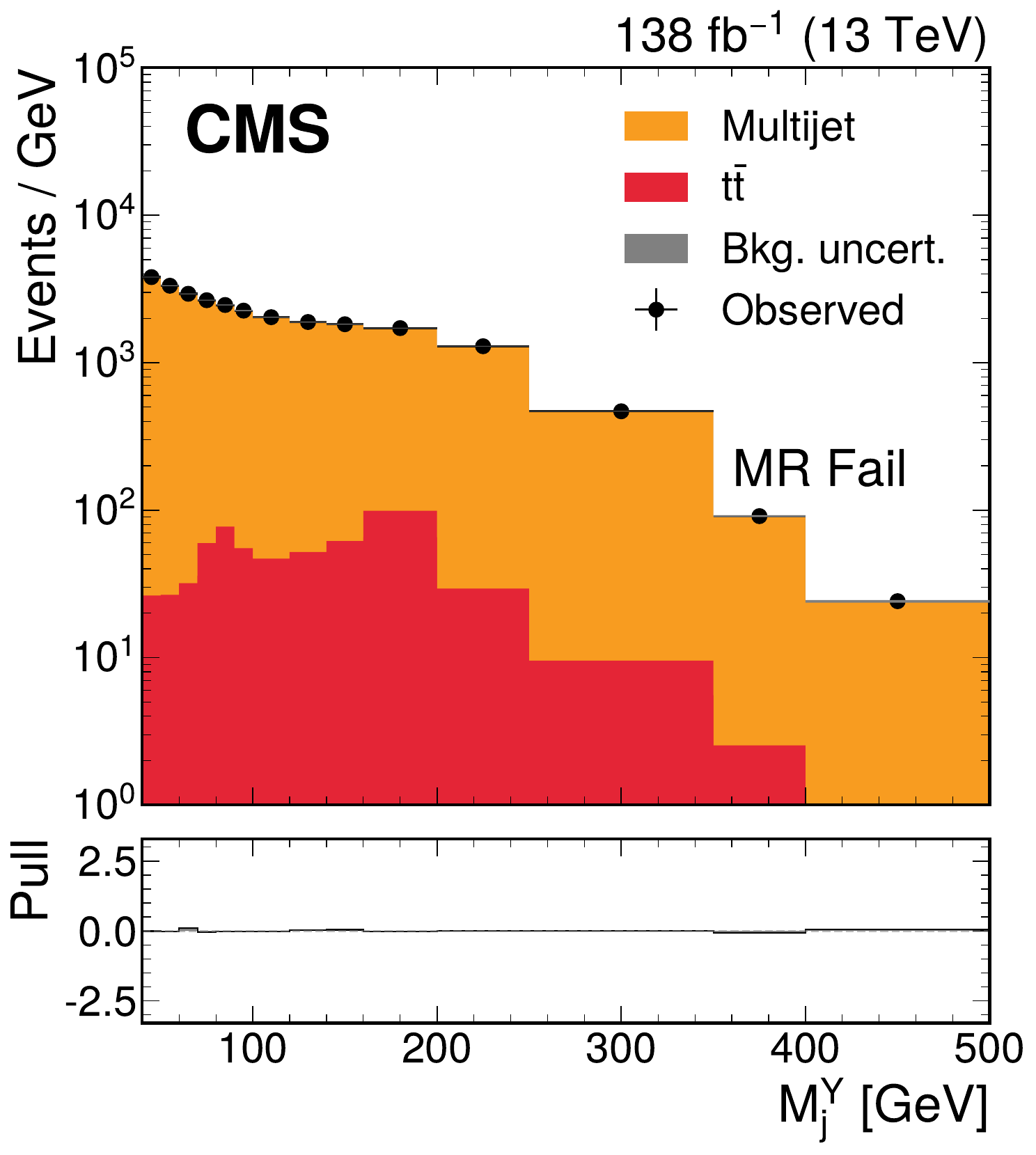}
  \caption{The \mjj (left) and \mjy (right) projections showing the number of observed events per \GeV (black markers) compared with the backgrounds estimated in the fit to the data (filled histograms) in the MR. Pass (upper) and Fail (lower) categories are shown. The expected contribution from the signal benchmark with $\MX = 2200\GeV$ and $\MY = 250\GeV$ is overlaid in the MR Pass, assuming a production cross section of 5\unit{fb}. The high level of agreement between the model and the data in the Fail region arises because this region is used to constrain the background estimate. The lower panels show the ``Pull'' defined as $(\text{observed events}{-}\text{expected events})/\sqrt{\smash[b]{\sigmaobs^{2} + \sigmaexp^{2}}}$, where \sigmaobs and \sigmaexp are the total uncertainties in the observation and the background estimation, respectively.}
  \label{fig:postfit}
\end{figure*}

Maximum likelihood fits are performed with the CMS statistical analysis tool \textsc{Combine}~\cite{CMS:2024onh} to estimate the background and extract the fitted signal. The fits are performed separately in the CR and MR. The F-test favors an \rpf polynomial order of $n=0\, (2)$ in the CR (MR) fit. Both fits pass the goodness-of-fit test, performed using the saturated model~\cite{saturated}. The $p$-value~\cite{Gross2010}, indicating the level of agreement with the background-only hypothesis and the observed data, is found to be 0.59 (0.26) in the CR (MR) fit. The postfit distributions from the CR and MR fits are shown in Figs.~\ref{fig:postfit_CR} and \ref{fig:postfit}, respectively.

The joint likelihoods of the signal-plus-background \mjj-\mjy distributions in the MR Pass and Fail regions are constructed for different signal hypotheses. The maximum likelihood fits are used to extract the observed signal strengths. Out of the 42 $\PX\to\PH(\bbbar)\PY(\PW\PW)$ mass points considered, the $\MX=1600\GeV$ and $\MY=90\GeV$ combination corresponds to the largest local significance of 2.1 standard deviations.

The same likelihoods are used to compute upper limits on the signal cross section with a modified frequentist approach, using the \CLs criterion~\cite{CLS2,CLS1} with the profile likelihood ratio used as the test statistic, in an asymptotic approximation~\cite{Cowan2011}. 

\begin{figure}[!htb]
  \centering 
  \includegraphics[width=0.49\textwidth]{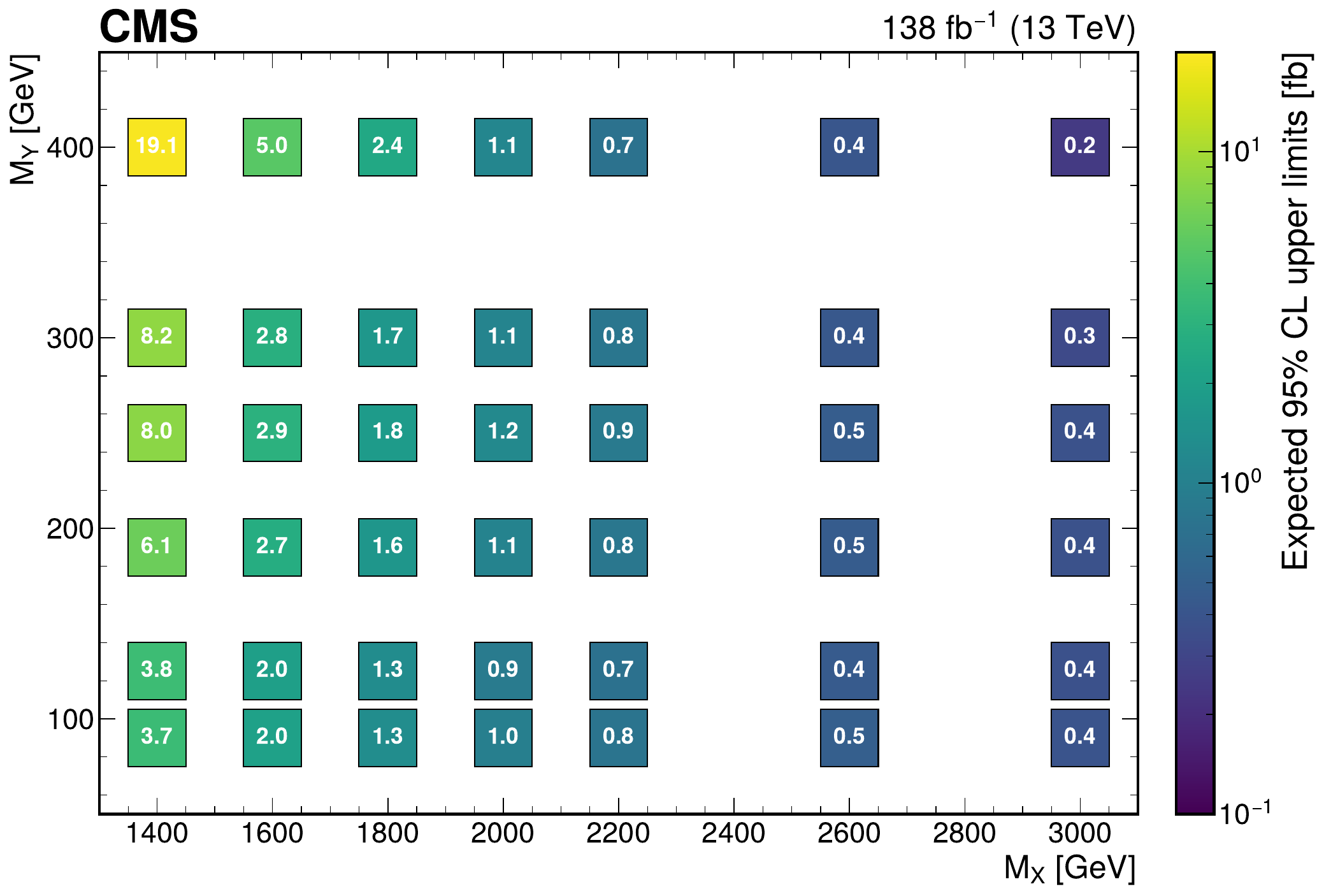}
  \includegraphics[width=0.49\textwidth]{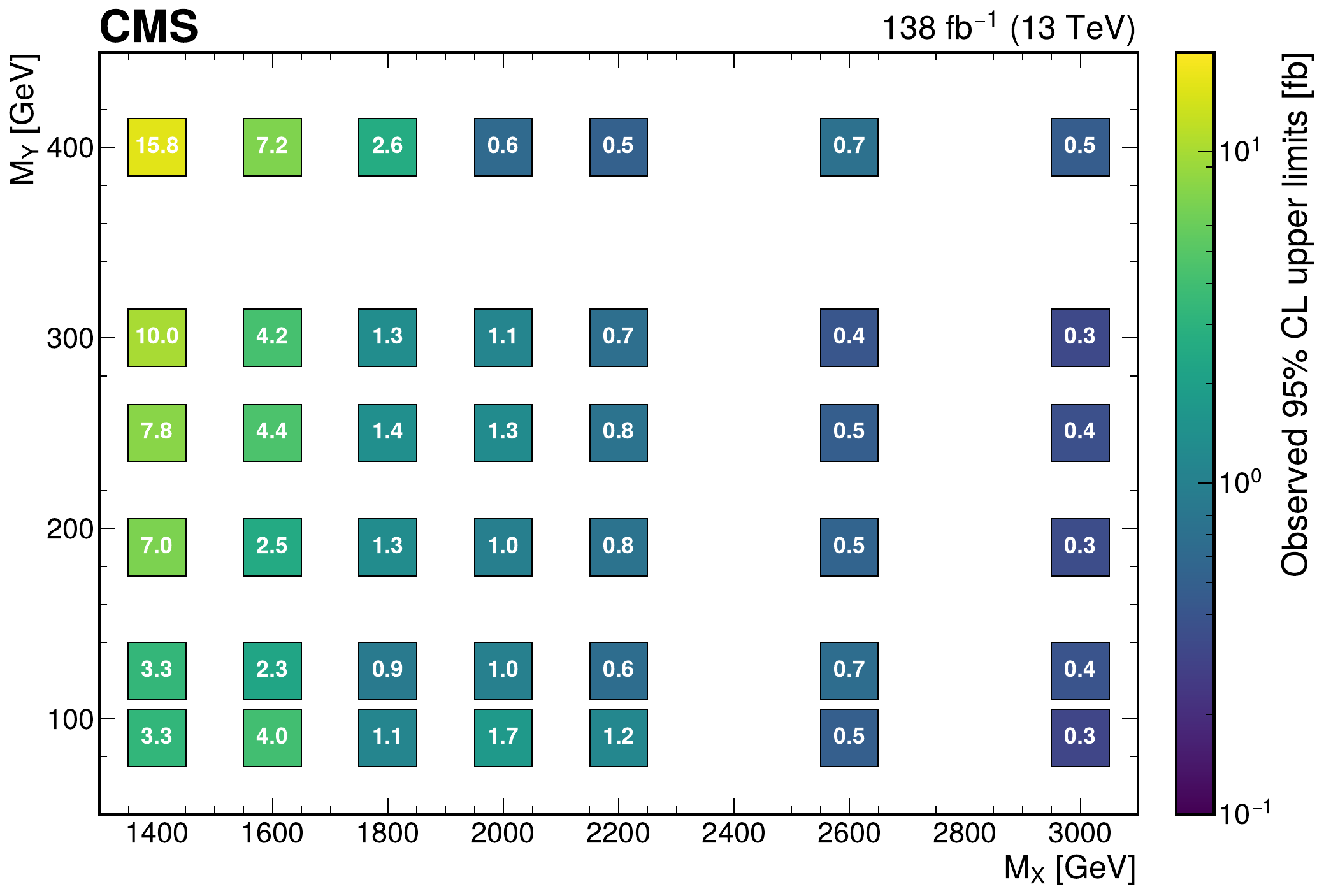}
  \caption{The expected (\cmsLeft) and observed (\cmsRight) 95\% confidence level upper limits on $\sigma(\Pp \to \XHY)\mathcal{B}(\PH\to\bbbar)\mathcal{B}(\PY\to\ww\to\qqqq)$ for different values of \MX and \MY.
    The limits have been evaluated in discrete steps corresponding to the centers of the boxes. The numbers in the boxes are given in fb.}
  \label{fig:lims2D}
\end{figure}

Figure~\ref{fig:lims2D} shows the expected and observed limits at 95\% confidence level (CL), assuming the \Yww decay to an all-hadronic final state. These limits range from 0.3--19.1\unit{fb}. In the upper left corner of the figures, where \MX is low and \MY is high, we begin to exit the Lorentz-boosted regime. This results in the \PY decay products being reconstructed as separate jets, reducing the efficiency of the offline selection and leading to higher upper limits.

\begin{figure*}[!htb]
  \centering 
  \includegraphics[width=0.49\textwidth]{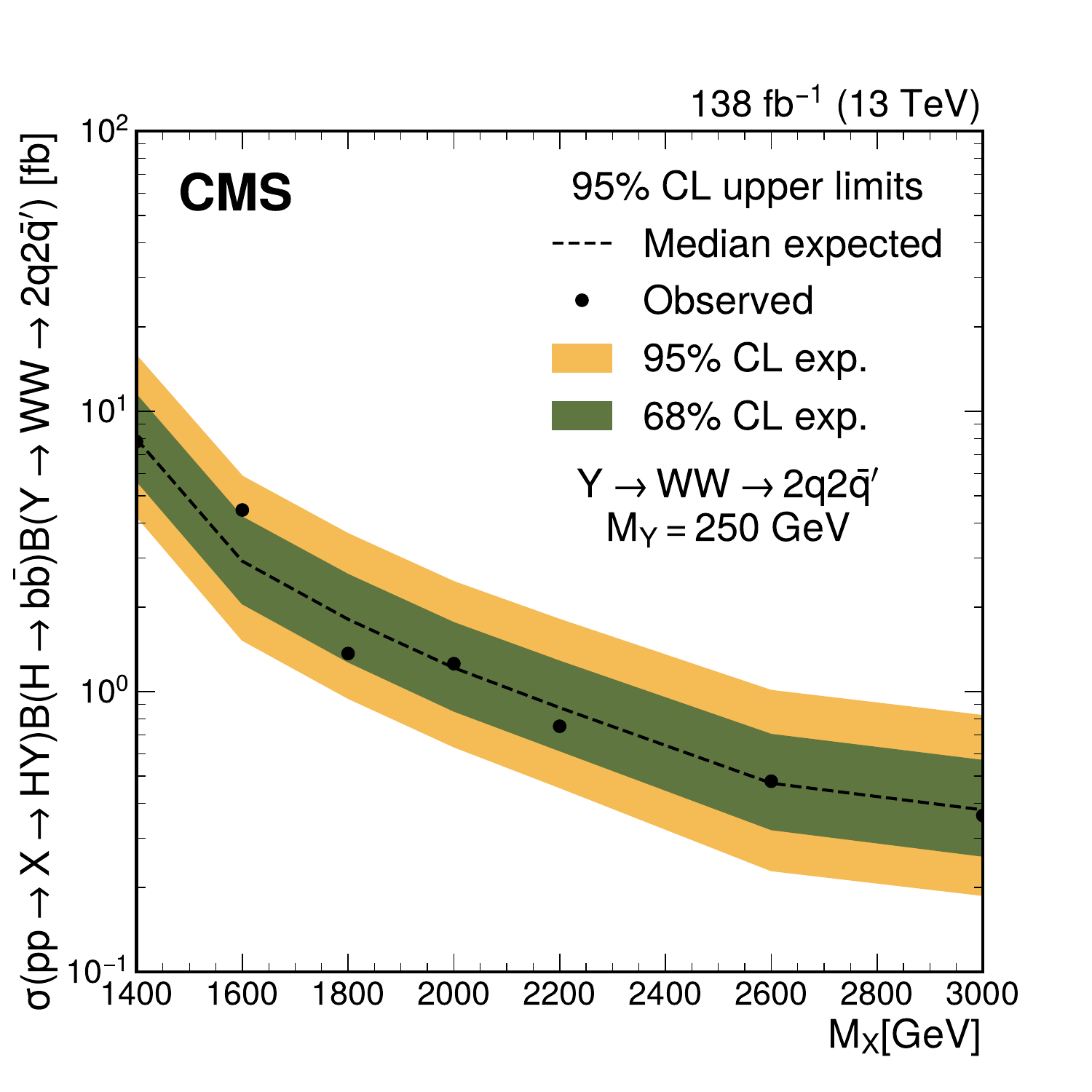}
  \includegraphics[width=0.49\textwidth]{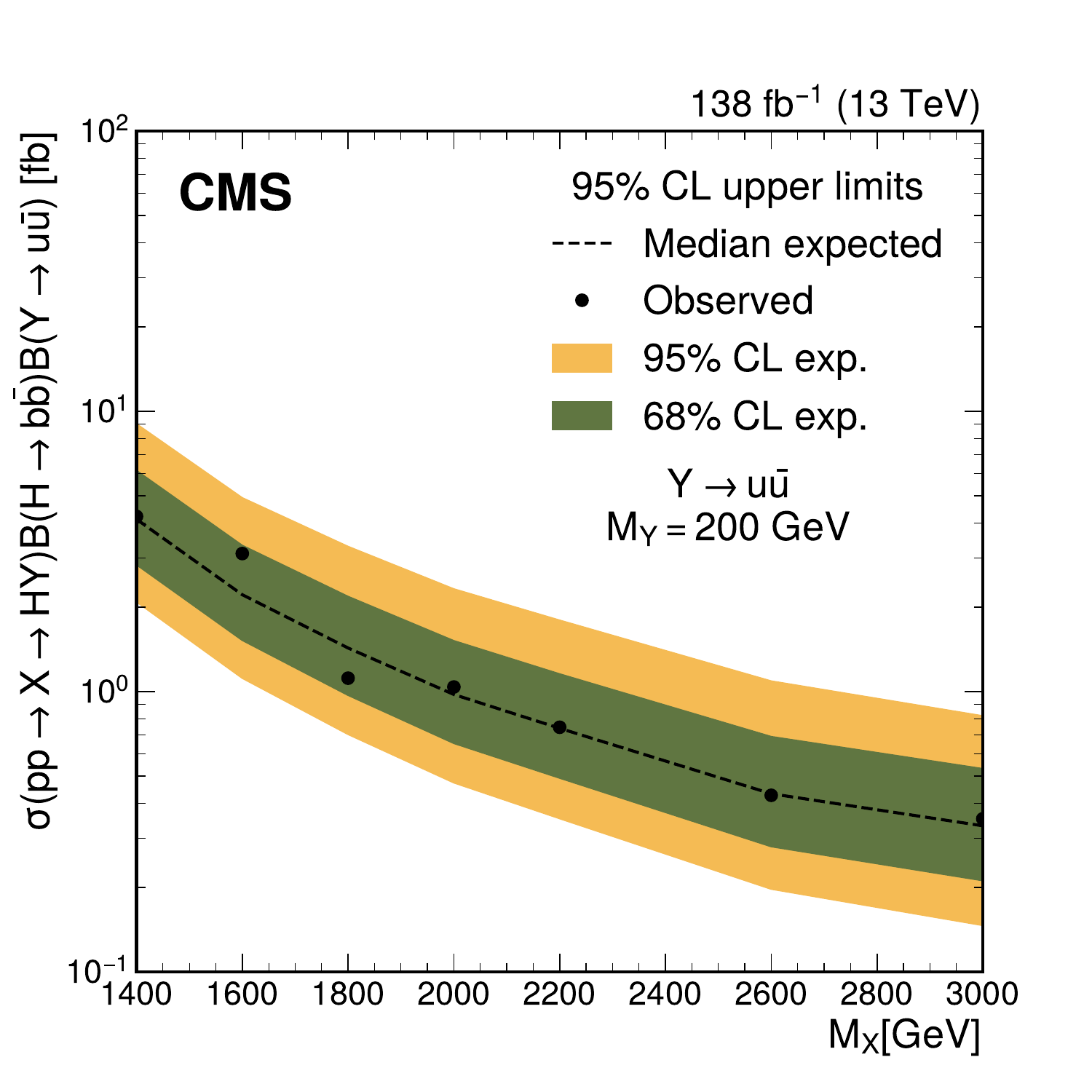}
  \includegraphics[width=0.49\textwidth]{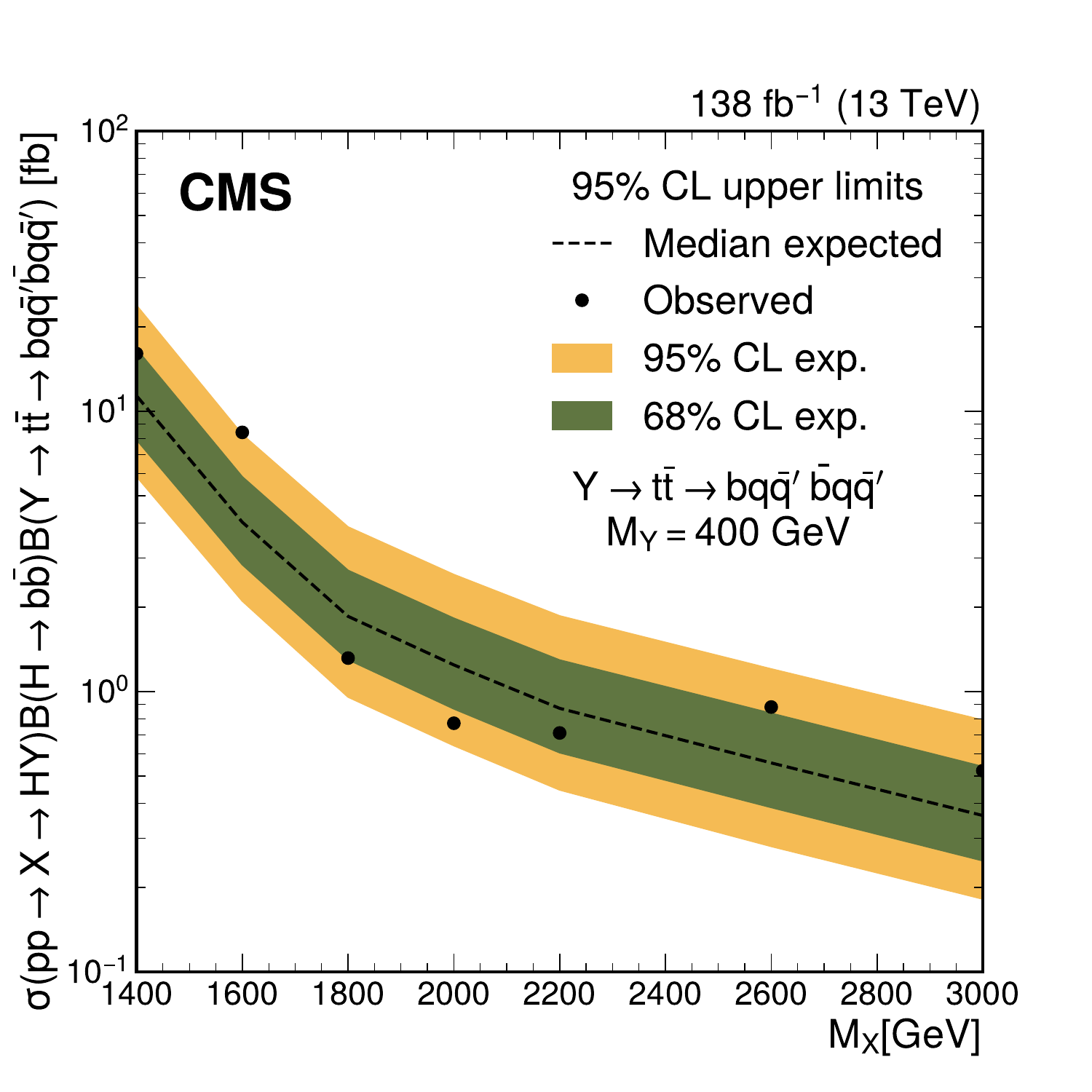}
  \includegraphics[width=0.49\textwidth]{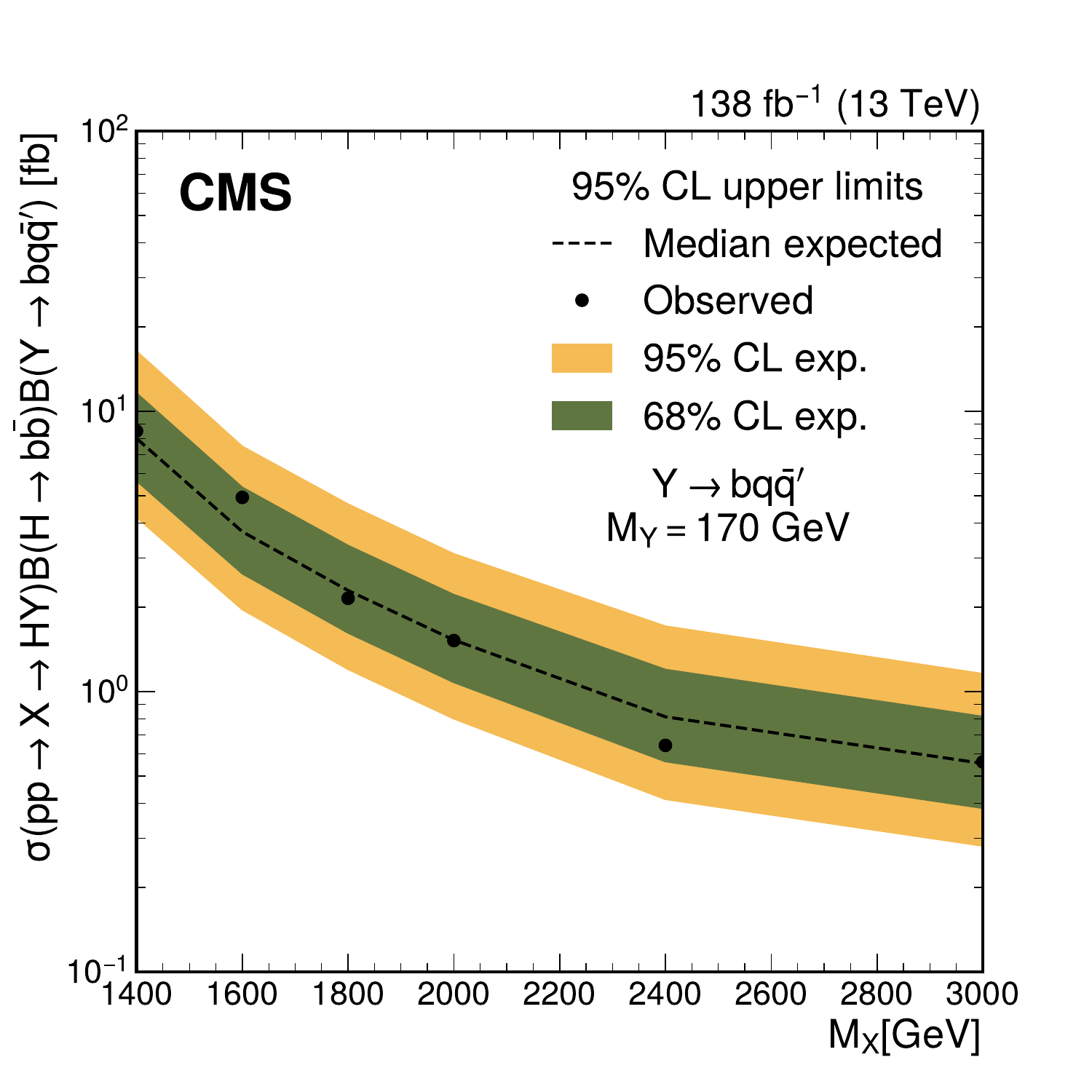}
  \caption{The median expected (dashed line) 95\% confidence level upper limits on the main and three alternative signal scenarios as a function of \MX. The inner (green) and outer (yellow) bands represent the regions containing 68\% and 95\%, respectively, of the distribution of limits expected under the background-only hypothesis. The solid dots indicate the observed limits, which are only evaluated at the simulated mass points.}
  \label{fig:limsAlt}
\end{figure*}

Comparable searches have been performed by the ATLAS Collaboration using both unsupervised~\cite{PhysRevLett.132.081801,ATLAS:XHbbYaj} and weakly supervised~\cite{2502.09770} anomaly detection techniques in two-body and dijet final states. The search in Ref.~\cite{PhysRevLett.132.081801} targets several two-body final states, while Ref.~\cite{ATLAS:XHbbYaj} specifically focuses on the $\PX \to \PH(\bbbar) \PY$ topology, where \PY decays to a fully hadronic state, using anomalous-jet tagging. These unsupervised searches explore broader ranges of both \MX and \MY than the analysis presented here, whereas the weakly supervised search~\cite{2502.09770} focuses on higher \MX regions. Finally, a targeted ATLAS search for $\PX\to\PH\PS$, with $\PH\to\gamma\gamma$ and $\PS\to\bbbar$ \cite{jhep11(2024)047}, explores complementary lower-mass regions with $\MX < 1\TeV$. Overall, these analyses provide complementary sensitivity across different mass ranges and final-state topologies.
The presented results can be compared to the CMS search for resonances decaying into two jets~\cite{case}, which reports expected model-independent limits on dijet resonances in the 50--600\unit{fb} range for \MX values between 1.8 and 3.0\TeV, depending on the method and \MX. The most sensitive method sets limits around 60\unit{fb} across this \MX range. Reference~\cite{case} also provides limits on a process that consists of a Kaluza--Klein excitation of a \PW boson (\WKK) decaying into a radion and a \PW boson~\cite{Agashe:2016rle,Agashe:2017wss}. The radion then decays into two \PW bosons, and all three \PW bosons undergo hadronic decays. The comparison to our results is motivated by the similarity in jet substructure: both analyses involve a four-prong jet from the \Yww decay and a two-prong jet originating from either \PH or \PW boson decay, where ``prong'' refers to the number of hard subjets within a large-radius jet. Although there are no directly overlapping mass points, we can compare the expected limits for a \WKK boson mass of 3000\GeV and a radion mass of 170\GeV (22.1\unit{fb}) with our limits for $\MX = 3000\GeV$ and $\MY = 200\GeV$ (0.4\unit{fb}). In all cases, our analysis sets more stringent limits, as expected, mostly because we require one of the jets to originate from \Hbb. While this requirement makes the results less general, it enables the use of \ParticleNet tagging, which provides most of the sensitivity in the analysis, with the autoencoder giving a modest additional improvement. Figure~\ref{fig:limsAlt} shows 95\% CL upper limits on the cross section for the three additional signal scenarios considered. Notably, the \Ytt scenario constitutes the first search performed in the Lorentz-boosted regime for this process. The differences in the limits among the scenarios arise mainly from their distinct kinematic properties and the varying background levels in the corresponding regions of phase space, both influenced by the different assumed \PY masses.

Tabulated results of this analysis are available on HEPData~\cite{hepdata}. The HepData entry also includes the trained AE weights used for the anomaly score evaluation.

\clearpage

\section{Summary}
\label{sec:Summary}

A search for beyond the standard model physics through the process where a resonant particle decays into a Higgs boson, \PH, and an additional particle \PY has been presented. The \PH subsequently decays into a bottom quark-antiquark pair, \bbbar, reconstructed as a large-radius jet in the Lorentz-boosted regime. The \PH candidate jets are tagged using the \ParticleNet algorithm that is designed to recognize jets originating from a decay of a massive particle into \bbbar. The identification of the second particle, \PY, is performed by computing the anomaly score of its candidate jet using an autoencoder, allowing the simultaneous search for multiple Y decay modes within a single analysis. This approach combines the targeted identification of \PH decays with model-independent anomaly detection technique, enabling a broad search for new physics. By combining the strong discrimination power of the \PH boson decay to \bbbar with a model-independent anomaly detection for the second particle, this analysis achieves enhanced sensitivity for scenarios involving a Higgs boson in the decay chain, while maintaining applicability across a range of such models.

The analysis considers four benchmark models. The main benchmark assumes \Yww, with further hadronic decays of the \PW bosons. It is simulated for \MX values within 1400--3000\GeV and \MY values within 90--400\GeV, covering 42 signal hypotheses. No significant excess above the standard model background expectations is observed. Upper limits on benchmark signal cross sections at 95\% confidence level are set on the main benchmark model in the 0.3--15.8\unit{fb} range. Additionally, exclusion limits are calculated for three alternative benchmark models, assuming \Yuu, \Ytt, or \Ybqq decay modes, for which the most stringent limits to date are achieved.

\begin{acknowledgments}
We congratulate our colleagues in the CERN accelerator departments for the excellent performance of the LHC and thank the technical and administrative staffs at CERN and at other CMS institutes for their contributions to the success of the CMS effort. In addition, we gratefully acknowledge the computing centers and personnel of the Worldwide LHC Computing Grid and other centers for delivering so effectively the computing infrastructure essential to our analyses. Finally, we acknowledge the enduring support for the construction and operation of the LHC, the CMS detector, and the supporting computing infrastructure provided by the following funding agencies: SC (Armenia), BMBWF and FWF (Austria); FNRS and FWO (Belgium); CNPq, CAPES, FAPERJ, FAPERGS, and FAPESP (Brazil); MES and BNSF (Bulgaria); CERN; CAS, MoST, and NSFC (China); MINCIENCIAS (Colombia); MSES and CSF (Croatia); RIF (Cyprus); SENESCYT (Ecuador); ERC PRG, TARISTU24-TK10 and MoER TK202 (Estonia); Academy of Finland, MEC, and HIP (Finland); CEA and CNRS/IN2P3 (France); SRNSF (Georgia); BMFTR, DFG, and HGF (Germany); GSRI (Greece); NKFIH (Hungary); DAE and DST (India); IPM (Iran); SFI (Ireland); INFN (Italy); MSIT and NRF (Republic of Korea); MES (Latvia); LMTLT (Lithuania); MOE and UM (Malaysia); BUAP, CINVESTAV, CONACYT, LNS, SEP, and UASLP-FAI (Mexico); MOS (Montenegro); MBIE (New Zealand); PAEC (Pakistan); MES, NSC, and NAWA (Poland); FCT (Portugal); MESTD (Serbia); MICIU/AEI and PCTI (Spain); MOSTR (Sri Lanka); Swiss Funding Agencies (Switzerland); MST (Taipei); MHESI and NSTDA (Thailand); TUBITAK and TENMAK (T\"{u}rkiye); NASU (Ukraine); STFC (United Kingdom); DOE and NSF (USA).

\hyphenation{Rachada-pisek} Individuals have received support from the Marie-Curie program and the European Research Council and Horizon 2020 Grant, contract Nos.\ 675440, 724704, 752730, 758316, 765710, 824093, 101115353, 101002207, 101001205, and COST Action CA16108 (European Union); the Leventis Foundation; the Alfred P.\ Sloan Foundation; the Alexander von Humboldt Foundation; the Science Committee, project no. 22rl-037 (Armenia); the Fonds pour la Formation \`a la Recherche dans l'Industrie et dans l'Agriculture (FRIA-Belgium); the Beijing Municipal Science \& Technology Commission, No. Z191100007219010, the Fundamental Research Funds for the Central Universities, the Ministry of Science and Technology of China under Grant No. 2023YFA1605804, and the Natural Science Foundation of China under Grant No. 12061141002 (China); the Ministry of Education, Youth and Sports (MEYS) of the Czech Republic; the Shota Rustaveli National Science Foundation, grant FR-22-985 (Georgia); the Deutsche Forschungsgemeinschaft (DFG), among others, under Germany's Excellence Strategy -- EXC 2121 ``Quantum Universe" -- 390833306, and under project number 400140256 - GRK2497; the Hellenic Foundation for Research and Innovation (HFRI), Project Number 2288 (Greece); the Hungarian Academy of Sciences, the New National Excellence Program - \'UNKP, the NKFIH research grants K 131991, K 133046, K 138136, K 143460, K 143477, K 146913, K 146914, K 147048, 2020-2.2.1-ED-2021-00181, TKP2021-NKTA-64, and 2021-4.1.2-NEMZ\_KI-2024-00036 (Hungary); the Council of Science and Industrial Research, India; ICSC -- National Research Center for High Performance Computing, Big Data and Quantum Computing, FAIR -- Future Artificial Intelligence Research, and CUP I53D23001070006 (Mission 4 Component 1), funded by the NextGenerationEU program (Italy); the Latvian Council of Science; the Ministry of Education and Science, project no. 2022/WK/14, and the National Science Center, contracts Opus 2021/41/B/ST2/01369, 2021/43/B/ST2/01552, 2023/49/B/ST2/03273, and the NAWA contract BPN/PPO/2021/1/00011 (Poland); the Funda\c{c}\~ao para a Ci\^encia e a Tecnologia, grant CEECIND/01334/2018 (Portugal); the National Priorities Research Program by Qatar National Research Fund; MICIU/AEI/10.13039/501100011033, ERDF/EU, "European Union NextGenerationEU/PRTR", and Programa Severo Ochoa del Principado de Asturias (Spain); the Chulalongkorn Academic into Its 2nd Century Project Advancement Project, and the National Science, Research and Innovation Fund via the Program Management Unit for Human Resources \& Institutional Development, Research and Innovation, grant B39G680009 (Thailand); the Kavli Foundation; the Nvidia Corporation; the SuperMicro Corporation; the Welch Foundation, contract C-1845; and the Weston Havens Foundation (USA).
\end{acknowledgments}

\bibliography{auto_generated}
\cleardoublepage \appendix\section{The CMS Collaboration \label{app:collab}}\begin{sloppypar}\hyphenpenalty=5000\widowpenalty=500\clubpenalty=5000
\cmsinstitute{Yerevan Physics Institute, Yerevan, Armenia}
{\tolerance=6000
A.~Hayrapetyan, V.~Makarenko\cmsorcid{0000-0002-8406-8605}, A.~Tumasyan\cmsAuthorMark{1}\cmsorcid{0009-0000-0684-6742}
\par}
\cmsinstitute{Institut f\"{u}r Hochenergiephysik, Vienna, Austria}
{\tolerance=6000
W.~Adam\cmsorcid{0000-0001-9099-4341}, J.W.~Andrejkovic, L.~Benato\cmsorcid{0000-0001-5135-7489}, T.~Bergauer\cmsorcid{0000-0002-5786-0293}, M.~Dragicevic\cmsorcid{0000-0003-1967-6783}, C.~Giordano, P.S.~Hussain\cmsorcid{0000-0002-4825-5278}, M.~Jeitler\cmsAuthorMark{2}\cmsorcid{0000-0002-5141-9560}, N.~Krammer\cmsorcid{0000-0002-0548-0985}, A.~Li\cmsorcid{0000-0002-4547-116X}, D.~Liko\cmsorcid{0000-0002-3380-473X}, M.~Matthewman, I.~Mikulec\cmsorcid{0000-0003-0385-2746}, J.~Schieck\cmsAuthorMark{2}\cmsorcid{0000-0002-1058-8093}, R.~Sch\"{o}fbeck\cmsAuthorMark{2}\cmsorcid{0000-0002-2332-8784}, D.~Schwarz\cmsorcid{0000-0002-3821-7331}, M.~Shooshtari, M.~Sonawane\cmsorcid{0000-0003-0510-7010}, W.~Waltenberger\cmsorcid{0000-0002-6215-7228}, C.-E.~Wulz\cmsAuthorMark{2}\cmsorcid{0000-0001-9226-5812}
% ...
\par}
\vskip\cmsinstskip
\dag:~Deceased\\
$^{1}$Also at Yerevan State University, Yerevan, Armenia\\
$^{2}$Also at TU Wien, Vienna, Austria\\
% ...
\end{sloppypar}
\end{document}